# GTPack: A Mathematica group theory package for application in solid-state physics and photonics

R. Matthias Geilhufe [1,*] and Wolfram Hergert [2]

[1]*Nordita, KTH Royal Institute of Technology and Stockholm University, Roslagstullsbacken 23, SE-106 91 Stockholm, Sweden*
[2]*Institute of Physics, Martin Luther University Halle-Wittenberg, Von-Seckendorff-Platz 1, 06120 Halle, Germany*

Correspondence*:
R. M. Geilhufe, W. Hergert
geilhufe@kth.se, wolfram.hergert@physik.uni-halle.de

## ABSTRACT

We present the Mathematica group theory package GTPack providing about 200 additional modules to the standard Mathematica language. The content ranges from basic group theory and representation theory to more applied methods like crystal field theory, tight-binding and plane-wave approaches capable for symmetry based studies in the fields of solid-state physics and photonics. GTPack is freely available via `http://GTPack.org`. The package is designed to be easily accessible by providing a complete Mathematica-style documentation, an optional input validation and an error strategy. We illustrate the basic framework of the package and show basic examples to present the functionality. Furthermore, we give a complete list of the implemented commands including references for algorithms within the supplementary material.

## INTRODUCTION

Symmetry and symmetry breaking are basic concepts of nature. Thus, arguments based on the symmetry of the considered system play a significant role within almost all branches of physics. Group theory represents the mathematical language to deal with symmetry, since all transformations that leave a physical system invariant (usually described in terms of transformations of algebraic or differential equations) naturally form a group. The application of group theory in physics has a long tradition ranging back to the beginning of the 20th century [1, 2]. Concentrating on solid-state and condensed matter physics, examples for the application of group theory can be found in the theory of the degeneracy of energy bands [3], color centers, $d^0$ magnetism and impurity states [4, 5, 6], optical transitions [7], phase transitions, atoms and molecules on surfaces [8], x-ray diffraction and crystallography in Fourier space [9], construction of effective low-energy excitation Hamiltonians [10], the classification of the superconducting states of matter [11, 12, 13, 14, 15], and, more recently, topological band theory [16, 17, 18, 19, 20] and topological quantum computation [21]. Due to the similarity of the underlying formalism, several concepts can be transferred to the field of photonics, for example, the band theory of photonic crystals [22, 23, 24], impurities and defect modes [25], and selection rules and uncoupled modes [26, 27].

In many cases, non-trivial results can be obtained from basic group theoretical information like the characters of the irreducible representations or the Clebsch-Gordan coefficients. In the past decades these information were tabulated in various books (e.g., [28, 29]). A comprehensive and widely used online group theory tool is provided by the Bilbao crystallographic server [30]. The work with printed group theory tables is not suitable for automation and the probability of copying and pasting misprints is high. However, modern computer algebra systems are capable to provide the same information, assumed that the necessary algorithms are implemented. Especially, the computer algebra system GAP (groups, algorithms and programming) [31] represents a powerful program to deal with computationally demanding questions in abstract algebra. Similarly, the computer algebra system Mathematica is well established within the research community. However, a stable group theory package designed for applications in solid-state physics and photonics is not included in the standard version.

The development of the Mathematica group theory package GTPack was designed to fill this gap. The functionality was planned to cover both, an application in active research and an application in university teaching. Therefore a main focus is the development of a user-friendly application, via self-explanatory names for new commands, a comprehensive documentation system and an optional input validation.

The aim of the paper is to report on the initial version of the Mathematica group theory package GTPack which is freely available for academic usage via `http://GTPack.org`. In the first part of the paper we introduce the main functionality and structure of the package. Afterwards we give general information about the implementation of the commands. In the last part we provide simple examples to illustrate the package. The supplementary material contains a full reference of all implemented modules. A more comprehensive guide about applied group theory in connection to GTPack can be found in Ref. [32].

## FUNCTIONALITY AND STRUCTURE

According to the functionality of the modules, GTPack is divided into various subpackages as illustrated in Fig. 1. In general, the subpackages can be assembled in three groups, "basic functionality", "structure" and "applications". Subpackages belonging to the class of "basic functionality" are *Auxiliary.m*, *Basic.m*, *Install.m* and *RepresentationTheory.m*. The package *Auxiliary.m* contains modules which are needed by GTPack or extend the general Mathematica language. Among others tesseral harmonics (real spherical harmonics) were added and also the Cartesian form of tesseral and spherical harmonics was implemented. Furthermore, the package contains modules for the handling with quaternions, the calculation of Gaunt coefficients, Dirac matrices or SU(2)-rotation matrices, to name but a few. The package *Basic.m* concentrates on general abstract group theory and, for example, provides modules for the calculation of classes, multiplication tables, left and right cosets and several logical commands to check for groups, Abelian groups, subgroups or invariant subgroups. Modules needed for basic representation theory are contained in the package *RepresentationTheory.m*. This package comprises the calculation of character tables, handling of irreducible representations, the calculation of Clebsch-Gordan coefficients, etc.. Modules for the installation of point and space groups or symmetry elements can be found in *Install.m*.

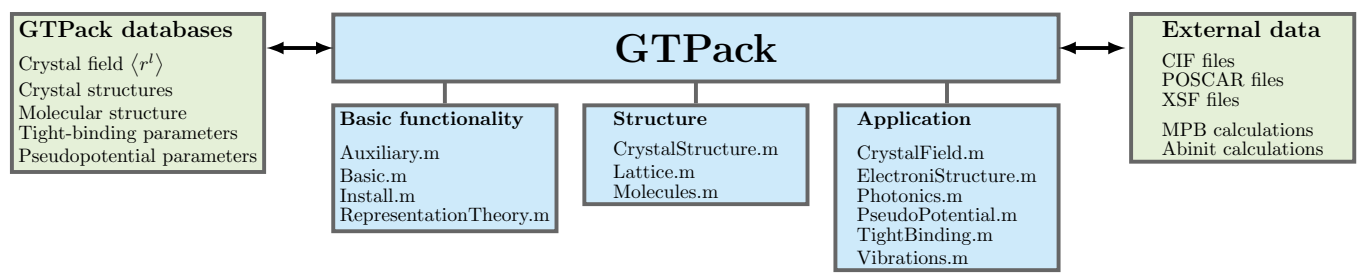

**Figure 1.** Subpackage structure of GTPack and interaction with external data.

The second class "structure" comprises of the packages *CrystalStructure.m*, *Lattice.m* and *Molecules.m*. Within *CrystalStructure.m*, modules for loading, saving and handling of crystal structures can be found. Modules with similar functionality but specialized on molecules are contained within *Molecules.m*. The construction and manipulation of atomic clusters as well as several commands for dealing with the reciprocal space are summarized in the package *Lattice.m*.

Next to basic group theory GTPack also contains subpackages for particular applications in solid state physics and photonics. The third class "Applications" contains the subpackages *CrystalField.m*, *ElectronicStructure.m*, *Photonics.m*, *PseudoPotential.m*, *TightBinding.m* and *Vibrations.m*. The crystal field package *CrystalField.m* is capable of automatically generating crystal field Hamiltonians. Furthermore, it contains the generation of standard operator equivalents like Stevens [?] and Buckmaster-Smith-Thornley [33] operators. GTPack also allows for electronic structure calculations for periodic systems, i.e. crystals, in the framework of the tight-binding and the pseudopotential approximation. Modules to construct tight-binding Hamiltonians are summarized in *TightBinding.m* and modules for the pseudopotential approximation in *PseudoPotential.m*. The calculation of band structures, density of states as well as the calculation of a Fermi surface is practically independent of the underlying model Hamiltonian. Therefore such commands are contained within the package *ElectronicStructure.m*. In the framework of plane-waves it is also possible to calculate the band structure of photonic crystals. The necessary modules for the construction of the master equation for various geometrical objects can be found in the package *Photonics.m*. Commands for the investigation of phonons or molecular vibrational modes are contained in *Vibrations.m*.

For various applications it is necessary to incorporate data, for example, tight-binding parameters, crystal field parameters and crystal structures. Therefore, GTPack contains several modules for the creation and handling of databases (cf. Fig. 1). Here, special file endings are used, such as *\*.parm* and *\*.struc*. Databases can be easily created, extended and modified by the user. Furthermore, GTPack includes commands for the interaction with external data formats and ab initio software. This concerns the import and export of structural data files *\*.cif* [34] and *\*.xsf* [35] and the output of the programs MIT Photonic Bands - MPB [36, 37]. For the future it is intended to implement similar modules for VASP [38] and abinit [39].

## IMPLEMENTATION

To distinguish new commands provided by GTPack from the standard Mathematica language and to prevent conflicts with new versions of Mathematica, all GTPack commands are starting with the characters *GT* (e.g. *GTInstallGroup*, *GTCharacterTable*, ...). Options are denoted with a suffix *GO* (e.g. *GOVerbose*, *GOIrepNotation*, ...). One of the main features of GTPack is the symbolic representation of symmetry elements. Symmetry elements for various standard axes (see Figure 2)) are predefined and the respective symbols, such as $C3z$ for a 3-fold rotation about the $z$-axis are protected. As a standard form they are displayed with subscripts, i.e., $C_{3z}$. Internally all symmetry elements are represented using matrices. The conversion between symbols and matrices can be done using *GTGetMatrix* and *GTGetSymbol*, respectively. Every module is implemented such that it is capable to handle lists of matrices in arbitrary representations. Depending on the application, GTPack uses a certain standard representation to provide a faithful matrix representation of point groups. Elements of ordinary point groups are represented by 3D rotation matrices of the group $O(3)$. In the special case of planar groups, the standard representation can be chosen to be $O(2)$. Symmetry elements within double groups are represented according to Damhus [40], where elements of groups not containing the inversion are represented by $SU(2)$ matrices and elements of groups containing the inversion are represented by matrices of the direct product group $SU(2) \otimes S$ with $S = \{1, -1\}$. Additionally, users can also define a group by providing a multiplication table. In this case GTPack automatically installs the provided elements within the multiplication table as new symmetry elements using permutation matrices as faithful representations.

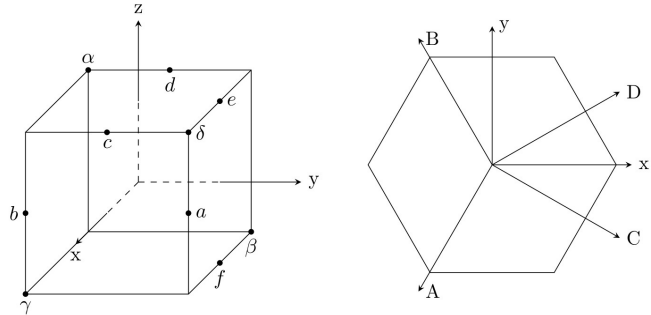

**Figure 2.** Rotation axes for symmetry elements in Schönflies notation.

Character tables are frequently needed and can be calculated using *GTCharacterTable*. The module uses the Burnside algorithm [41], which is a reasonable choice for relatively small groups, such as point groups. Representation matrices can be generated using *GTGetIrep*, where the algorithm of Flodmark and Blokker is implemented [42]. Clebsch-Gordan coefficients, which are necessary to generate the basis of a direct product representation, can be calculated using *GTClebschGordanCoefficients*. Here, the algorithm of van Den Broek and Cornwell is implemented [43]. To calculate band structures of solids, GTPack supports a plane-wave basis [44, 45] and a tight-binding method in the two- and three-center form [46, 47]. The Master equation for photonic crystals is constructed by *GTPhMaster* as described in Ref. [24].

## INSTALLATION OF GTPACK

GTPack is installed similarly to all other Mathematica packages. After downloading and decompressing GTPack, the content of the package has to be copied to the application folder of Mathematica within the respective base directory. Here, the user can choose between making the package available for all users of the computer or only for her- or himself. If the package should be available for all users of the computer, the corresponding folder to copy to can be found by opening a Mathematica notebook and typing $BaseDirectory. If the package should be available exclusively for the current user of the computer the respective base directory can be found by typing $UserBaseDirectory. According to the path of the base or user base directory (called $dir in the following), the folder containing GTPack needs to be copied to the directory $dir\Applications. Afterwards, the package and the documentation are available. The package itself can be loaded within a Mathematica notebook by typing Needs["GroupTheory`"].

# EXAMPLES

## Installation of groups and character table

In the first example, the installation of point groups and the calculation of character tables will be shown. Within the example, the point group $T$ is considered. Using GTPack, the point group is installed with the command *GTInstallGroup* as shown in Fig. 3. The output is a list of symmetry elements, where each element is given in symbolic form. In total $T$ contains 12 elements, where the symmetry elements are denoted using the Schönflies notation [32], where $C_{na}$ denotes a rotation about the angle $2\pi/n$ about the rotation axis $\vec{a}$. The implemented standard axes are shown in Figure 2. Additional rotation axes can be installed using *GTInstallAxis*. Each symbol can be transformed into a rotation matrix using *GTGetMatrix*. A character table for a point group is installed using *GTCharacterTable*.

The command applies the Burnside algorithm for the calculation of the character table [41]. Within the command several options can be specified optionally, such as an input validation (*GOFast*), a

```
In[2]:= Needs["GroupTheory`"]
In[3]:= group = GTInstallGroup[T]
        The standard representation has changed to O(3)
Out[3]= {Ee, C_{2z}, C_{2y}, C_{2x}, C_{3β}, C_{3γ}, C_{3γ}^{-1}, C_{3α}^{-1}, C_{3β}^{-1}, C_{3δ}^{-1}, C_{3α}, C_{3δ}}
In[4]:= GTCharacterTable[group, GOIrepNotation → "Mulliken", GOReality → True];
```

|   | Ee | 3 $C_{2z}$ | 4 $C_{3\beta}$ | 4 $C_{3\gamma}^{-1}$ | Reality |
|---|---|---|---|---|---|
| A | 1 | 1 | 1 | 1 | potentially real |
| $E^2$ | 1 | 1 | $-\frac{1}{2}i(-i+\sqrt{3})$ | $\frac{1}{2}i(i+\sqrt{3})$ | essentially complex |
| $E^1$ | 1 | 1 | $\frac{1}{2}i(i+\sqrt{3})$ | $-\frac{1}{2}i(-i+\sqrt{3})$ | essentially complex |
| T | 3 | -1 | 0 | 0 | potentially real |

$C_1 = \{Ee\}$

$C_2 = \{C_{2z}, C_{2y}, C_{2x}\}$

$C_3 = \{C_{3\beta}, C_{3\gamma}, C_{3\alpha}, C_{3\delta}\}$

$C_4 = \{C_{3\gamma}^{-1}, C_{3\alpha}^{-1}, C_{3\beta}^{-1}, C_{3\delta}^{-1}\}$

```
In[5]:= doublegroup = GTInstallGroup[T, GORepresentation → "SU(2)"]
        The standard representation has changed to SU(2)
Out[5]= {Ee, Eē, C_{3δ}, C_{3γ}^{-1}, C_{3β}^{-1}, C_{3α}, C_{3α}^{-1}, C_{3β}, C_{3γ}, C_{3δ}^{-1}, C_{2y},
         C_{2x}, C̄_{2x}, C̄_{2y}, C̄_{2z}, C_{2z}, C̄_{3δ}^{-1}, C̄_{3γ}, C̄_{3β}, C̄_{3α}^{-1}, C̄_{3α}, C̄_{3δ}, C̄_{3γ}^{-1}, C̄_{3δ}}
In[6]:= GTCharacterTable[doublegroup];
```

|   | Ee | 4 $C_{3\delta}$ | 4 $C_{3\gamma}^{-1}$ | 4 $C_{3\delta}^{-1}$ | 4 $C_{3\gamma}$ | 6 $C_{2y}$ | Eē |
|---|---|---|---|---|---|---|---|
| $\Gamma^1$ | 1 | 1 | 1 | 1 | 1 | 1 | 1 |
| $\Gamma^2$ | 1 | $-(-1)^{1/3}$ | $\frac{1}{2}i(i+\sqrt{3})$ | $\frac{1}{2}i(i+\sqrt{3})$ | $-(-1)^{1/3}$ | 1 | 1 |
| $\Gamma^3$ | 1 | $\frac{1}{2}i(i+\sqrt{3})$ | $-(-1)^{1/3}$ | $-(-1)^{1/3}$ | $\frac{1}{2}i(i+\sqrt{3})$ | 1 | 1 |
| $\Gamma^4$ | 2 | -1 | -1 | 1 | 1 | 0 | -2 |
| $\Gamma^5$ | 2 | $\frac{1}{2}(1-i\sqrt{3})$ | $\frac{1}{2}(1+i\sqrt{3})$ | $-(-1)^{1/3}$ | $\frac{1}{2}i(i+\sqrt{3})$ | 0 | -2 |
| $\Gamma^6$ | 2 | $\frac{1}{2}(1+i\sqrt{3})$ | $\frac{1}{2}(1-i\sqrt{3})$ | $\frac{1}{2}i(i+\sqrt{3})$ | $-(-1)^{1/3}$ | 0 | -2 |
| $\Gamma^7$ | 3 | 0 | 0 | 0 | 0 | -1 | 3 |

$C_1 = \{Ee\}$

$C_2 = \{C̄_{3\delta}, C̄_{3\alpha}, C̄_{3\beta}, C̄_{3\gamma}\}$

$C_3 = \{C_{3\gamma}^{-1}, C_{3\beta}^{-1}, C_{3\alpha}^{-1}, C_{3\delta}^{-1}\}$

$C_4 = \{C_{3\delta}^{-1}, C_{3\alpha}^{-1}, C_{3\beta}^{-1}, C_{3\gamma}^{-1}\}$

$C_5 = \{C_{3\gamma}, C_{3\beta}, C_{3\alpha}, C_{3\delta}\}$

$C_6 = \{C_{2y}, C_{2x}, C̄_{2x}, C̄_{2y}, C̄_{2z}, C_{2z}\}$

$C_7 = \{Eē\}$

**Figure 3.** Installation of groups and double groups and calculation of the character tables using the commands *GTInstallGroup* and *GTCharacterTable*.

control of the printed output (*GOVerbose*), or the choice of notation for the irreducible representations (*GOIrepNotation*). For example, possible options for the names of irreducible representations are: the Mulliken notation [48, 49], which is widely used in chemistry and spectroscopy; the notation according to Bouckaert, Smoluchowski and Wigner [3], which is usually used in connection to band structure calculations; and a simple index notation, which is denoted by Bethe notation. To determine additional degeneracies of energy levels due to time-reversal symmetry, the reality of irreducible representations plays a central role [50]. Given a point group $\mathcal{G}$, a representation $\Gamma$ of $\mathcal{G}$ is called: potentially real, if $\Gamma$ is equivalent to a real representation and $\Gamma \sim \Gamma^*$; pseudo-real, if $\Gamma$ is not equivalent to a real representation, but $\Gamma \sim \Gamma^*$; and essentially complex, if $\Gamma \nsim \Gamma^*$. This property can be determined by means of the equation of Frobenius and Schur [50], given by

$$\frac{1}{g}\sum_{T\in\mathcal{G}} \chi(T^2) = \begin{cases} 1 & \text{if } \Gamma \text{ is potentially real} \\ 0 & \text{if } \Gamma \text{ is essentially complex} \\ -1 & \text{if } \Gamma \text{ is pseudo-real} \end{cases}. \quad (1)$$

Equation (1) can be evaluated using *GTReality*, or during the calculation of the character table by specifying the option *GOReality*. In the second part of the example in Figure 3 the character table is calculated for the double group of $T$. The double group is installed by specifying the option *GORepresentation* within the command *GTInstallGroup* and choosing $SU(2)$ as standard representation. The additional symbols of the double group elements are denoted with an overline. Instead of four, the double group of $T$ has seven classes and irreducible representations. The additional classes are classified by the theorem of Opechowski [51, 52]. The respective extra representations of the double group can be determined using *GTExtraRepresentations*.

**Crystal-field splitting**

Crystal field theory represents a semi-empirical approach to describe localized states in an atomic or crystallographic surrounding. The crystallographic surrounding or crystal field is described in terms of a small perturbation $V_{\mathrm{cr}}(\vec{r})$ leading to the total Hamiltonian

$$\left[-\frac{\hbar^2}{2m}\nabla^2 + V(\vec{r}) + V_{\mathrm{cr}}(\vec{r})\right]\psi(\vec{r}) = E\psi(\vec{r}). \quad (2)$$

The crystal field itself can be expanded in terms of spherical harmonics $Y_m^l$ [32],

$$V_{\mathrm{cr}}(\vec{r}) = \sum_l \sum_{m=-l}^{l} r^l A_{l,m} Y_m^l(\theta,\phi), \quad (3)$$

or more generally in terms of crystal field operators $\hat{O}_m^l$,

$$\hat{V}_{\mathrm{cr}} = \sum_l \sum_{m=-l}^{l} B_{l,m} \hat{O}_m^l. \quad (4)$$

GTPack contains various modules to calculate matrix elements $\left\langle jm_1|\hat{O}_m^l|jm_2\right\rangle$ for operator equivalents, such as spherical harmonics, Buckmaster-Smith-Thornley operators [33], and Stevens operators

[? ]. The respective GTPack commands are given by *GTGauntCoefficients*, *GTBSTOperator*, and *GTStevensOperator*.

Depending on the underlying symmetry of the system some of the expansion coefficients $A_{l,m}$ or $B_{l,m}$ are zero. Given a symmetry group $\mathcal{G}$, then the Hamiltonian of the system and with that the crystal field expansion as to be invariant under the application of the projection operator of the identity representation,

$$\hat{\mathcal{P}}^1 V_{\text{cr}}(r, \theta, \phi) = V_{\text{cr}}(r, \theta, \phi). \tag{5}$$

However, for any proper coordinate transformation $\hat{P}(T)$ corresponding to an element $T \in \mathcal{G}$ the spherical harmonics (and similarly crystal field operators $\hat{O}_m^l$) transform as

$$\hat{P}(T) Y_l^m = \sum_{m'=-l}^{l} D_{m'm}^l(T) Y_l^{m'}, \tag{6}$$

where $D_{m'm}^l$ denotes the Wigner-D function. For improper coordinate transformations (inversion, reflections, etc.) a factor of $(-1)^l$ has to be taken into account. Evaluating equation (5) by using the

```
We consider the point group O_h and calculate the crystal field Hamiltonian.
In[*]:= Ohgroup = GTInstallGroup[Oh, GOVerbose → "False"];
       vcr = GTCrystalField[Ohgroup, 4]
Out[*]= A[0, 0] Y[0, 0] + r^4 A[4, 4] Y[4, -4] + Sqrt[14/5] r^4 A[4, 4] Y[4, 0] + r^4 A[4, 4] Y[4, 4]

According to linear perturbation theory the matrix A^l_{m1,m2} = ⟨Y_l^{m1}, V_cr Y_l^{m2}⟩ is calculated. This can be
simplified by using GTGauntCoefficient.
In[*]:= Amat = vcr /. Y[l_, m_] :>
         Table[(-1)^m1 GTGauntCoefficient[2, -m1, l, m, 2, m2], {m1, -2, 2}, {m2, -2, 2}];

In the following the eigenvalues for an octahedral, a cubic and a tetrahedral crystal field are calculated
and the parameters Dq[oct] and ϵ[oct] are introduced.
In[*]:= rule = {q r^4 → 24 π Dq[oct], q → 2/3 π ϵ[oct]};

Octahedron
In[*]:= Roct = {{R, 0, 0}, {-R, 0, 0}, {0, R, 0}, {0, -R, 0}, {0, 0, R}, {0, 0, -R}} /. R → 1;
In[*]:= qoct = Table[q, {i, 1, 6}];
In[*]:= Expand[Eigenvalues[Amat]] /.
         A[l_, m_] -> GTCrystalFieldParameter[l, m, Roct, qoct] /. rule
Out[*]= {-4 Dq[oct] + ϵ[oct], -4 Dq[oct] + ϵ[oct],
       -4 Dq[oct] + ϵ[oct], 6 Dq[oct] + ϵ[oct], 6 Dq[oct] + ϵ[oct]}

Cube
In[*]:= Rcube = {{R, R, R}, {-R, R, R}, {-R, -R, R}, {R, -R, R},
                {R, R, -R}, {-R, R, -R}, {-R, -R, -R}, {R, -R, -R}} / Sqrt[3] /. R → 1;
In[*]:= qcube = Table[q, {i, 1, 8}];
In[*]:= Expand[Eigenvalues[Amat]] /.
         A[l_, m_] -> GTCrystalFieldParameter[l, m, Rcube, qcube] /. rule
Out[*]= {32 Dq[oct]/9 + 4 ϵ[oct]/3, 32 Dq[oct]/9 + 4 ϵ[oct]/3,
       32 Dq[oct]/9 + 4 ϵ[oct]/3, -16 Dq[oct]/3 + 4 ϵ[oct]/3, -16 Dq[oct]/3 + 4 ϵ[oct]/3}

Tetrahedron
In[*]:= Rtet = {{R, -R, R}, {-R, R, R}, {-R, -R, -R}, {R, R, -R}} / Sqrt[3] /. R → 1;
In[*]:= qtet = Table[q, {i, 1, 4}];
In[*]:= Expand[Eigenvalues[Amat]] /.
         A[l_, m_] -> GTCrystalFieldParameter[l, m, Rtet, qtet] /. rule
Out[*]= {16 Dq[oct]/9 + 2 ϵ[oct]/3, 16 Dq[oct]/9 + 2 ϵ[oct]/3,
       16 Dq[oct]/9 + 2 ϵ[oct]/3, -8 Dq[oct]/3 + 2 ϵ[oct]/3, -8 Dq[oct]/3 + 2 ϵ[oct]/3}
```

**Figure 4.** Crystal field expansion and level splitting for a localized $d$-electron in an octahedral, a cubic and a tetrahedral crystal field.

expansion (3) and transformation behavior (6) leads to the under determined equation system

$$A_l^m = \frac{1}{g} \sum_{T \in \mathcal{G}} \sum_{m'=-l}^{l} D_{mm'}^l(T) A_l^{m'}. \tag{7}$$

From equation (7) it can be concluded which coefficients depend on each other and which coefficients vanish. The final symmetry adapted crystal field expansion can be calculated using GTPack by means of *GTCrystalField*. Figure 4 illustrates an example for the level splitting for a single $d$-electron in an octahedral, a cubic and a tetrahedral crystal field. The underlying point groups are $O_h$ for the cubic and the octahedral case as well as $T_d$ for the tetrahedral case. However, it turns out that both groups lead to the same crystal field expansion. First, the point group $O_h$ is installed using *GTInstallGroup*. Afterwards the crystal field expansion is calculated using *GTCrystalField* up to a cutoff value of $2l$. As we discuss $d$-electrons, the expansion is truncated after $l = 4$. Next, we define the surrounding field in terms of the nearest neighbors and create lists containing their positions and ionic charges. The splitting is calculated from the eigenvalues of the matrix elements over radial wave functions $\psi_m^l(\vec{r}) = R^l(r) Y_m^l(\theta, \phi)$, as

$$V_{\text{cr}} = \sum_{l'=0}^{2l} \sum_{m=-l'}^{l'} \left\langle r^{l'} \right\rangle A_{l',m} \left\langle lm_1 | l'm | lm_2 \right\rangle, \tag{8}$$

where

$$\left\langle r^{l'} \right\rangle = \int \mathrm{d}r \, r^2 r^l R^l(r)^2, \tag{9}$$

and

$$\left\langle lm_1 | l'm | lm_2 \right\rangle = \int \mathrm{d}\Omega Y_{m_1}^{l*}(\Omega) Y_m^{l'}(\Omega) Y_{m_2}^l(\Omega). \tag{10}$$

The latter integral is called Gaunt coefficient and can be calculated by means of GTPack using *GTGauntCoefficient*. The values for $\left\langle r^{l'} \right\rangle$ are materials specific and can be calculated, e.g., in the framework of the density functional theory [53]. GTPack provides commands to store and load explicit values from a database. However, in this example we introduce the generic parameters

$$\epsilon[\text{oct}] = \frac{3q \left\langle r^2 \right\rangle}{2\pi}, \tag{11}$$

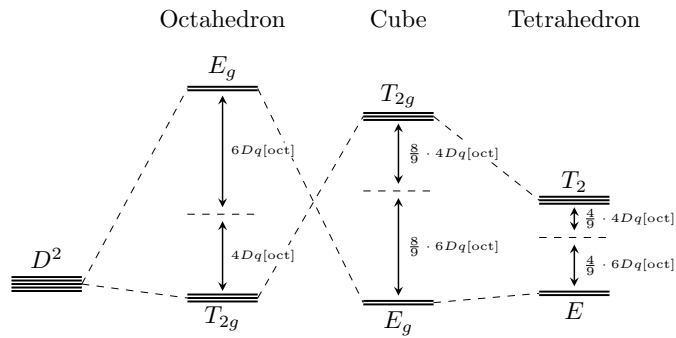

**Figure 5.** The level splitting for a localized $d$-electron into $E_g$ ($E$) and $T_{2g}$ ($T_2$) levels in an octahedral, a cubic and a tetrahedral crystal field.

and
$$\text{Dq[oct]} = \frac{q \left\langle r^4 \right\rangle}{24\pi}. \tag{12}$$

As can be verified from the Mathematica example in Figure 4, in all three cases a two-fold and a three-fold degenerate level can be found. In the cubic and octahedral case the two-fold degenerate state corresponds to the irreducible representation $E_g$ and the three-fold degenerate level corresponds to the irreducible representation $T_{2g}$. For the tetrahedron the inversion symmetry is broken and the states are denoted by $E$ and $T_2$, respectively. The final result is plotted in Figure 5.

### Tight-binding bandstructure of graphene

Due to the occurrence of Dirac nodes within the band structure and the resulting properties, graphene counts as one of the most studied materials to date [54, 55]. Within the following example the calculation of the band structure using the tight-binding approximation will be illustrated in the framework of GTPack. In Fig. 6 the structure and the construction of the tight-binding Hamiltonian is shown. At first, the structure is given as a list in the standard GTPack format. The list contains the name of the structure and a prototype, four different names for the space group (Pearson symbol, Strukturbericht designation, international notation, and space group number), the lattice, and the sites containing the atom name and the atom position. Note that the first five information are optional and not important for the generation of the tight-binding model. GTPack provides modules to import structures, e.g., in the cif-format using *GTImportCIF*. The structural information can be plotted using the command *GTPlotStructure2D*. For the construction of the tight-binding model it is necessary to construct a real space cluster. This cluster is reordered into different shells corresponding to nearest neighbor, next nearest neighbor, next next nearest neighbor interactions, etc.. The respective commands to do so are *GTCluster* and *GTShells*. From the information of the shells, the tight-binding Hamiltonian is constructed for $s$- and $p$-electrons. The zero and non-zero entries within the tight-binding Hamiltonian are illustrated using *GTHamiltonianPlot*. As can be seen within the plot, the $p_z$-orbitals do not hybridize with all the other orbitals. Hence, those can be considered to form a smaller tight-binding Hamiltonian of dimension $2 \times 2$. Setting up and solving the reduced $2 \times 2$ tight-binding Hamiltonian is shown in Fig. 7. The high-symmetry path within the Brillouin zone ($K'$, $\Gamma$, $M$, $K$) is generated using the command *GTBZPath*. The points $K$ and $K'$ denote the corners of the hexagonal Brillouin zone, $M$ points to the middle of an edge and $\Gamma$ is the Brillouin zone center. The band structure itself is calculated and plotted using *GTBandStructure*.

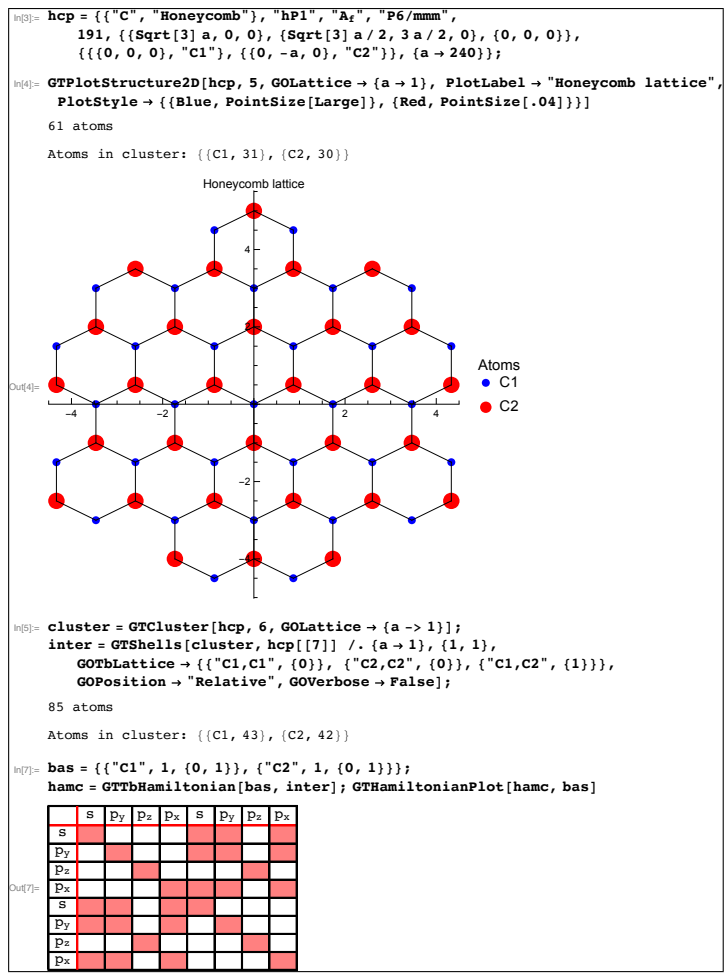

**Figure 6.** Structural information and construction of the tight-binding Hamiltonian for graphene.

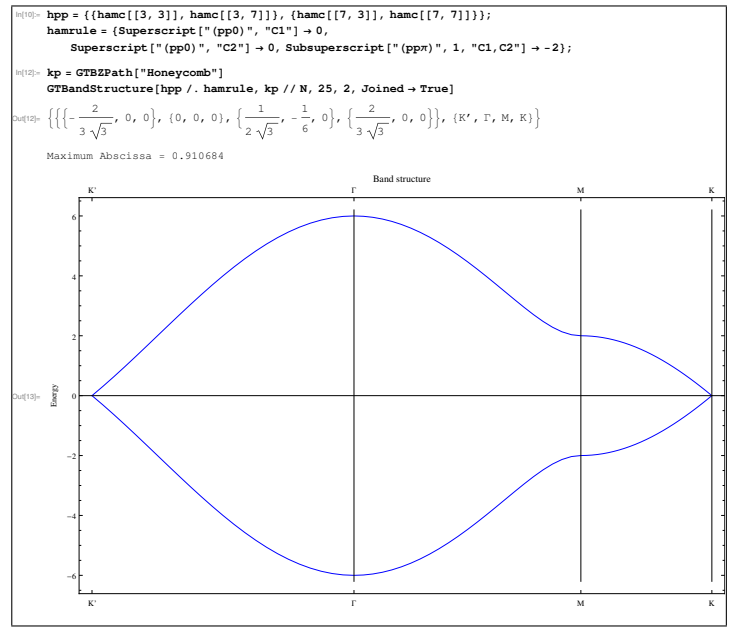

**Figure 7.** Band structure calculation for graphene.

## ANALYZING PHOTONIC BAND STRUCTURES

Photonic crystals represent the optical analogues of ordinary crystals, where light is traveling through a periodic dielectric. Potential optical band gaps, i.e., forbidden frequencies where photons are not allowed to travel through the medium have motivated research towards for various applications replacing ordinary electronics and information technology [56]. Recently, photonic crystals have been discussed with respect to nodal points [57] and topological states in periodic and quasi periodic arrangements [58, 59, 60]. A group theory classification of the allowed modes yields to additional information, e.g., with respect to uncoupled bands [27]. For two-dimensional photonic crystals, the vectorial Maxwell's equations can be transformed into two sets of independent equations for two modes [24], which are referred to as transversal

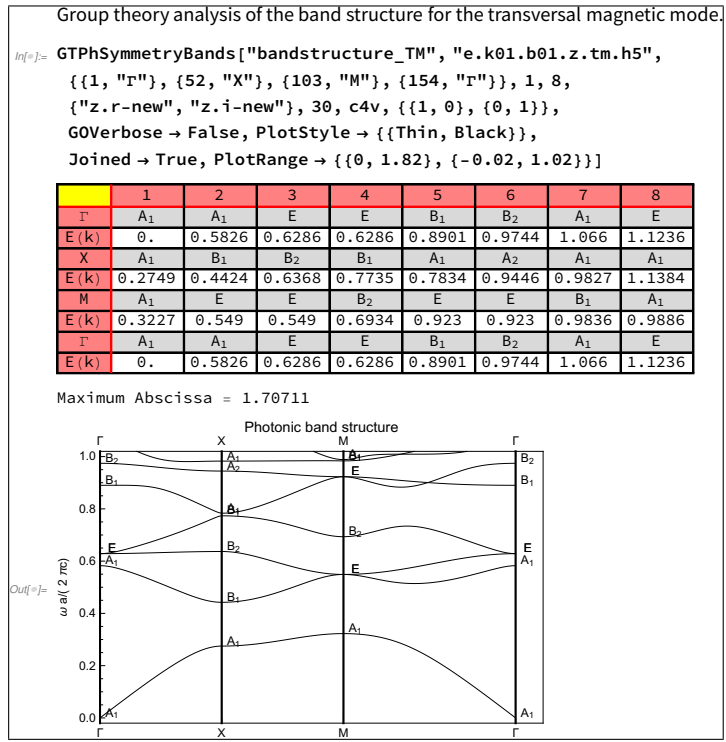

**Figure 8.** Photonic band structure for a two-dimensional square lattice of alumina rods in air calculated using MPB and analyzed with GTPack.

magnetic (TM) and transversal electric (TE) modes. The resulting master equations are given by

$$-\frac{1}{\varepsilon(\vec{r}_\parallel)}\left\{\frac{\partial^2}{\partial x^2}+\frac{\partial^2}{\partial y^2}\right\}\mathrm{E}_z(\vec{r}_\parallel)=\frac{\omega^2}{c^2}\mathrm{E}_z(\vec{r}_\parallel)\,, \tag{13}$$

$$-\left\{\frac{\partial}{\partial x}\frac{1}{\varepsilon(\vec{r}_\parallel)}\frac{\partial}{\partial x}+\frac{\partial}{\partial y}\frac{1}{\varepsilon(\vec{r}_\parallel)}\frac{\partial}{\partial y}\right\}\mathrm{H}_z(\vec{r}_\parallel)=\frac{\omega^2}{c^2}\mathrm{H}_z(\vec{r}_\parallel)\,. \tag{14}$$

Here, the vector $\vec{r}_\parallel$ denotes a vector in the $xy$ plane. For the solution of the masters equation, a plane-wave approach can be applied which transforms the Masters equations into an eigenvalue problem. Such an approach is implemented in GTPack, but also within the code MPB [37]. GTPack can be applied to analyze photonic band structure calculations performed with MPB, as will be shown in the following. We consider a two-dimensional photonic crystal with a square lattice made of circular alumina rods ($\varepsilon = 8.9$) in air. The radius of the rods is given by $R = 0.2a$ ($a$-lattice constant). This corresponds to a filling factor of $\gamma = 0.126$. The photonic band structure was calculated using MPB incorporating a tolerance of $10^{-7}$. The calculated band structure can be loaded, plotted and analyzed automatically using the command *GTPhSymmetryBands* as shown in Figure 8 for the transversal magnetic mode. The underlying point group is $C_{4v}$, which has four one-dimensional irreducible representation ($A_1$, $A_2$, $B_1$, $B_2$) and one two-dimensional irreducible representation ($E$). As can be seen in Figure 8, there is a spectral gap between the first and the second band. Degeneracies in the band structure correspond to the dimensions of the corresponding irreducible representation. Therefore, all single bands are associated to the one-dimensional irreducible representations. However, for example, for the second and the third band, a two-fold degeneracy can be found at the $M$ point which corresponds to the two-dimensional irreducible representation $E$. To revise this point more clearly, the specific fields corresponding to the modes can be imported into the Mathematica notebook using *GTPhMPBFields*. Afterwards, the field is analyzed by applying the character projection operator for each irreducible representation. As expected, only the application of the operator corresponding to $E$ shows non-zero results, as can be seen in Figure 9.

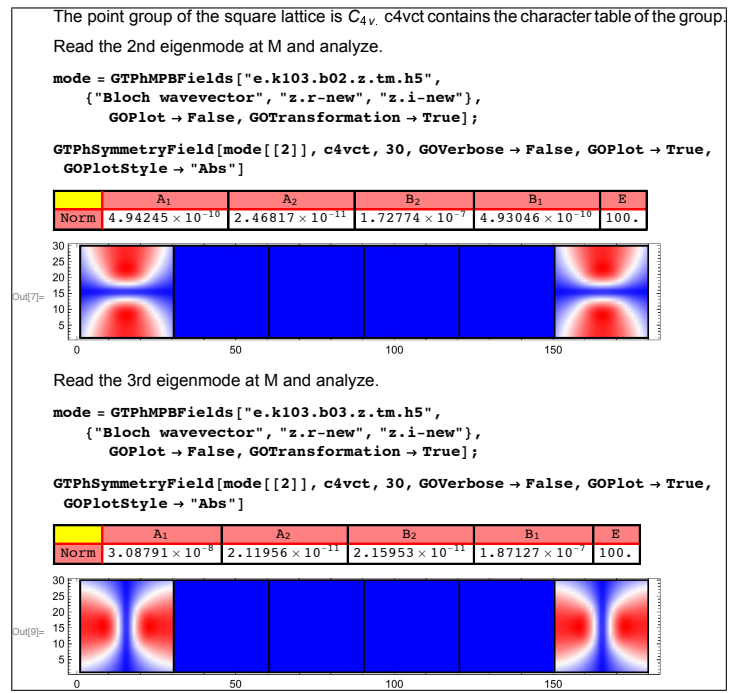

**Figure 9.** Application of the character projection operator to the field corresponding to the degeneracy between the second and third band of the transversal magnetic mode at $M$.

## ADVANTAGES AND LIMITATIONS

As an additional package to the standard Mathematica framework, GTPack is embedded into the Mathematica framework. The new modules provided within GTPack are designed for application in solid state physics and photonics and use notations which are common in these research communities. Having a programming framework at hand comes with the advantage of easy automation which is in contrast to recently published group theory tables. Most of the provided modules are kept general and can be applied in connection to any set of matrices forming a group. This allows for applications way beyond the provided point and space group setup. As usual, the package is constructed in a modular form and can be extended easily.

GTPack is under ongoing development and therefore comes with limitations in the first version. Among others, these comprise of the calculation of character tables for space groups and groups containing anti-unitary symmetry elements. The implemented modules for the calculation of photonic band structures currently do not reach the same performance as specialized numeric implementations such as MPB. Thus, export and import modules connect MPB with GTPack. Additionally, modules to export analytically generated tight-binding Hamiltonians to Fortran help to generate more efficient numeric codes. An extension to a parallel implementation for the usage of the cluster version of Mathematica is currently not planned. Concerning the symmetry analysis, irreducible representations can currently only be associated to the calculated bands if symmorphic space groups are taken into account. An extension for non-symmorphic groups is under development.

## CONCLUSION

We presented the Mathematica group theory package GTPack together with four basic examples. The package contains about 200 additional commands dealing with basic group theory and representation theory and providing tools for applications in solid state physics and photonics. The package itself is structured into several subpackages. In connection to external databases it is possible to load, change and save data, like structural information or parameters for electronic and photonic band structure calculations. The package works externally with a symbolic representation of symmetry elements. Internally, matrices are used. GTPack can be obtained online via the web page `http://GTPack.org`.

## ACKNOWLEDGMENTS

The authors acknowledge the support of Sebastian Schenk, especially with respect to his help in building up a documentation within the Mathematica framework. Furthermore, we are grateful for discussions and contributions from Markus Däne, Christian Matyssek, Stefan Thomas, Martin Hoffmann, and Arthur Ernst. This publication was funded by the German Research Foundation within the Collaborative Research Centre 762 (projects A4 and B1). Additionally, RMG acknowledges funding from the European Research Council under the European Union's Seventh Framework Program (FP/2207-2013)/ERC Grant Agreement No. DM-321031.

---

## SUPPLEMENTARY MATERIAL

The supplementary material comprises of a complete list of implemented modules. Each table from Table 1 to Table 13 contains the respective modules of one subpackage (see main text Fig. 1). The tables contain the name of the module, a short and general explanation and in some cases (when appropriate) a reference to the implemented algorithm or to further information.

### Auxiliary.m

Table 1    Modules of the subpackage *Auxiliary.m*. The package contains additional modules which were needed for GTPack but which were not covered in the standard Mathematica language.

| Name | Functionality | Ref. |
|---|---|---|
| **GTBlueRed** | defines the color scheme for tables printed within GTPack | |
| **GTCartesianSphericalHarmonicY** | calculates spherical harmonics $Y_l^m$ in Cartesian form | |
| **GTCartesianTesseralHarmonicY** | calculates tesseral harmonics $S_l^m$ in Cartesian form | |
| **GTClusterFilter** | removes certain atoms from a given cluster | |
| **GTCompactStore** | stores sparse matrices in compact form | |
| **GTDiracMatrix** | gives the Dirac matrices $\hat{\Gamma}_i$ in various representations | |
| **GTEulerAnglesQ** | checks if an input is a set of Euler angles | |
| **GTFermiSurfaceXSF** | calculates the band structure and stores results to XSF file | |
| **GTGauntCoefficient** | calculates the Gaunt coefficients | |
| **GTGroupConnection** | plots a graph illustrating the connection between several groups | |
| **GTGroupHierarchy** | plots a graph illustrating all subgroups of a group | |
| **GTNeighborPlot** | present information about input data to construct adjacency matrices | |
| **GTPointGroups** | illustrates the subgroup relationships of the 32 point groups | |
| **GTQAbs** | calculates the absolute value of a quaternion | |
| **GTQConjugate** | calculates the conjugate of a quaternion | |
| **GTQInverse** | calculates the inverse of a quaternion | |
| **GTQMultiplication** | multiplies two quaternions | |
| **GTQPolar** | calculates the polar angle of a quaternion | |
| **GTQuaternionQ** | checks if an input is a quaternion | |
| **GTReadFromFile** | reads an object from a file | |
| **GTSetTableColors** | defines a color function based on the MPB color code | |
| **GTSU2Matrix** | calculates a SU(2) matrix from a given angle and 3D vector | |
| **GTSymbolQ** | checks if an input is a GTPack symbol | |
| **GTTesseralHarmonicY** | calculates tesseral harmonics $S_l^m$ | [61] |
| **GTWriteToFile** | writes an object to a file | |

**Basic.m**

Table 2   Modules of the subpackage ***Basic.m***. The package contains modules for basic group theory, like calculation of classes, multiplication tables, cosets, etc..

| Name | Functionality | Ref. |
|---|---|---|
| **GTAbelianQ** | checks if list of elements forms an Abelian group | |
| **GTAllSymbols** | gives a listed of all currently installed symmetry elements | |
| **GTCenter** | calculates the center of a group | |
| **GTClasses** | calculates the classes of a group | |
| **GTClassMult** | calculates the class multiplication of two classes | |
| **GTClassMultTable** | calculates a class multiplication table | |
| **GTConjugacyClass** | constructs all conjugate subgroups for a given group and subgroup | |
| **GTConjugateElement** | calculates the conjugate element of an element | |
| **GTCyclicQ** | checks if list of elements forms a cyclic group | |
| **GTGenerators** | estimates a set of generators of a group | |
| **GTGetEulerAngles** | gives the Euler angles associated with a symmetry element | |
| **GTGetMatrix** | gives the matrix associated with a symmetry element | |
| **GTGetQuaternion** | gives the quaternion associated with a symmetry element | |
| **GTGetRotationMatrix** | gives the 3D rotation matrix associated with a symmetry element | |
| **GTGetSU2Matrix** | gives the SU(2) matrix associated with a symmetry element | |
| **GTGetSubGroups** | calculates the subgroups of a given group | |
| **GTGetSymbol** | gives the symbol associated with a symmetry element | |
| **GTgmat** | group multiplication | |
| **GTGroupOrder** | calculates the order of a group | |

| | |
|---|---|
| **GTGroupQ** | checks if list of elements forms a group |
| **GTInverseElement** | calculates the inverse of a symmetry element |
| **GTInvSubGroupQ** | checks if one group is an invariant subgroup of another one |
| **GTInvSubGroups** | calculates the invariant subgroups of a given group |
| **GTLeftCosets** | constructs left cosets for a given group and subgroup |
| **GTMagnetic** | logical module to allow for magnetic groups |
| **GTMagneticQ** | checks if magnetic groups are considered |
| **GTMultTable** | calculates the multiplication table of a group |
| **GTNormalizer** | calculates the normalizer of a group |
| **GTOrderOfElement** | calculates the order of a symmetry element |
| **GTProductGroup** | constructs the product group of two groups |
| **GTProductGroupQ** | checks if a group is a product of two given groups |
| **GTQuotientGroup** | constructs the quotient group of two groups |
| **GTQuotientGroupQ** | checks if two groups can form a quotient group |
| **GTRightCosets** | constructs right cosets for a given group and subgroup |
| **GTSelfAdjointQ** | checks if a symmetry element is self-adjoint |
| **GTSetMultiplication** | multiplies the elements of two sets with each other |
| **GTSubGroupQ** | checks if one group is a subgroup of another one |
| **GTTransformation** | applies a transformation operator to a given vector |
| **GTTransformationOperator** | applies a transformation operator to a given vector |
| **GTType** | gives a symmetry element in space group form |
| **GTWhichInput** | checks if the input is a symbol, matrix, quaternion or set of angles |
| **GTWhichOutput** | transforms an output to a symbol, matrix, quaternion or set of angles |

**CrystalField.m**

Table 3   Modules of the subpackage *CrystalField.m*. The package contains modules needed within crystal field theory, like calculation of the crystal field Hamiltonian, crystal field operators and crystal field parameters.

| Name | Functionality | Ref. |
|---|---|---|
| **GTBSTOperator** | calculates a matrix for Buckmaster-Smith-Thornley operator equivalent | [33] |
| **GTBSTOperatorElement** | calculates matrix elements for Buckmaster-Smith-Thornley operator equivalents | [33] |
| **GTCFDatabaseInfo** | gives a list of stored $\langle r^l \rangle$ expectation values | |
| **GTCFDatabaseRetrieve** | retrieves stored $\langle r^l \rangle$ expectation values from a file | |
| **GTCFDatabaseUpdate** | adds $\langle r^l \rangle$ expectation values to a file | |
| **GTCrystalField** | calculates an effective crystal field Hamiltonian | |
| **GTCrystalFieldParameter** | calculates crystal field parameters in the point-charge model | |
| **GTCrystalFieldSplitting** | calculates the decomposition of irreducible representations within a subgroup | |
| **GTStevensOperator** | calculates a matrix for Stevens operator equivalent | [62] |
| **GTStevensOperatorElement** | calculates matrix elements for Stevens operator equivalents | [62] |
| **GTStevensTheta** | gives multiplet prefactors for the crystal field expansion | [?] |

## CrystalStructure.m

Table 4   Modules of the subpackage *CrystalStructure.m*. The package contains modules for the handling of crystal structures and crystal structure databases.

| Name | Functionality | Ref. |
|---|---|---|
| **GTAllStructures** | prints a list of currently installed structures | |
| **GTBravaisLattice** | gives lattice vectors for a specified crystal system | |
| **GTBuckyBall** | gives a cluster of carbon atoms in buckminsterfullerene structure | |
| **GTClusterManipulate** | allows for manipulation of atomic clusters | |
| **GTCrystalData** | provides information about crystal structures | |
| **GTCrystalSystem** | gives a list of point groups for a specified crystal system | |
| **GTExportXSF** | exports structures to XCRYSDEN format | |
| **GTGetStructure** | gives an installed crystal structure | |
| **GTGroupNotation** | change notation between Schönflies and Hermann-Mauguin | |
| **GTImportCIF** | imports crystal structure information from CIF files | [34] |
| **GTInstallStructure** | installs new crystal structures | |
| **GTLatticeVectors** | constructs lattice vectors from lattice parameters and angles | |
| **GTLoadStructures** | load stored crystal structures | |
| **GTPlotCluster** | plots a cluster of atoms | |
| **GTPlotStructure** | plots an installed crystal structure | |
| **GTPlotStructure2D** | plots an installed 2D crystal structure | |
| **GTSaveStructures** | saves a set of installed crystal structures | |
| **GTShowSymmetryElements** | plots symmetry elements of a point group | |
| **GTSpaceGroups** | gives information about nomenclature of the 230 space groups | |
| **GTTubeParameters** | calculates the important geometric properties of single wall carbon nanotubes | |
| **GTTubeStructure** | constructs a $(n, m)$ single wall carbon nanotube. | |

## ElectronicStructure.m

Table 5   Modules of the subpackage *ElectronicStructure.m*. The package contains general modules for electronic structure calculations like calculation of band structures or density of states from various effective Hamiltonians.

| Name | Functionality | Ref. |
|---|---|---|
| **GTBands** | calculates the energy bands for an effective Hamiltonian | |
| **GTBandsPlot** | plots a calculated band structure | |
| **GTBandStructure** | calculates and plots the band structure for an effective Hamiltonian | |
| **GTCompatibility** | calculates the compatibility relations of two point groups | |
| **GTDensityOfStates** | calculates the density of states for an effective Hamiltonian | |
| **GTDensityOfStatesPlot** | plots a calculated density of states | |

| | |
|---|---|
| **GTDensityOfStatesRS** | calculates the density of states for an effective Hamiltonian in real space |
| **GTFermiSurface** | calculates the Fermi-surface |

## Install.m

Table 6  Modules of the subpackage *Install.m*. The package contains modules for the installation of groups and symmetry elements.

| Name | Functionality | Ref. |
|---|---|---|
| **GTChangeRepresentation** | changes the standard representation used by GTPack (SO(3), SO(2), SU(2), ...) | |
| **GTGroupFromGenerators** | installs a finite group from a set of generators | |
| **GTInstallAxis** | installs symmetry elements for a given rotation axis | |
| **GTInstallGroup** | installs point and space groups | |
| **GTReinstallAxes** | changes between active and passive convention for symmetry elements | |
| **GTTableToGroup** | install a permutation group from the multiplication table | |
| **GTWhichAxes** | informs if active or passive convention for symmetry elements is used | |
| **GTWhichRepresentation** | gives the currently used standard representation (SO(3), SO(2), SU(2), ...) | |

## Lattice.m

Table 7  Modules of the subpackage *Lattice.m*. The package contains modules connected to crystallographic lattices, like calculation of reciprocal lattice vectors, illustration of Wigner-Seitz cells and Brillouin zones and the construction of atomic clusters.

| Name | Functionality | Ref. |
|---|---|---|
| **GTAdjacencyMatrix** | calculates an adjacency matrix for a cluster of atoms | |
| **GTBZLines** | generates a set of $\vec{k}$-points for a Brillouin zone path | |
| **GTBZMPBPointMesh** | exports a set of $\vec{k}$-points in MPB data format | [36, 37] |
| **GTBZPath** | generates a standard $\vec{k}$-path in the Brillouin zone | |
| **GTBZPointMesh** | gives a $\vec{k}$-point mesh in the irreducible part of the Brillouin zone | |
| **GTCluster** | constructs a spherical cluster of atoms | |
| **GTGroupOfK** | calculates the group of the wave-vector $\vec{k}$ | |
| **GTLatCluster** | constructs a spherical cluster of lattice points | |
| **GTLatShells** | separates a cluster of lattice points into distinct shells | |
| **GTReciprocalBasis** | calculates the reciprocal basis | |
| **GTShells** | separates a cluster of atoms into distinct shells | |
| **GTStarOfK** | calculates the star of the wave-vector $\vec{k}$ | |
| **GTVoronoiCell** | plots the Brillouin zone or Wigner-Seitz cell | |

## Molecules.m

Table 8   Modules of the subpackage *Molecules.m*. The package contains modules for the handling of molecules and molecule structure databases.

| Name | Functionality | Ref. |
|---|---|---|
| **GTMolChemicalData** | gives molecular data from the Mathematica database in GTPack form | |
| **GTMolDatabaseInfo** | gives information of molecular data stored within a database | |
| **GTMolDatabaseUpdate** | add molecular information to a database | |
| **GTMolGetMolecule** | gives information of a particular molecule stored within a database | |
| **GTMolPermutationRep** | gives a representation of a group in terms of permutation matrices | |
| **GTMolToCluster** | transforms a molecule from a data set into a cluster of atoms | |

## Photonics.m

Table 9   Modules of the subpackage *Photonics.m*. The package contains modules for the study of photonic crystals.

| Name | Functionality | Ref. |
|---|---|---|
| **GTPhBandsObjects** | | |
| **GTPhCuboid** | gives the structure factor for a cuboid | |
| **GTPhDCObjects** | calculates the Fourier transform of the inverse permittivity of a set of geometric objects | |
| **GTPhDCPixel** | calculates the Fourier transform of the inverse permittivity defined by a pixel map | |
| **GTPhDielectric** | calculates the Fourier transform of $\epsilon^{-1}(\vec{r})$ | |
| **GTPhEllipticRod** | gives the structure factor of an elliptic rod | |
| **GTPhFields** | calculates the electro magnetic field in a photonic crystal | |
| **GTPhMaster** | constructs the master equation for a photonic crystal structure | [24] |
| **GTPhMasterObjects** | constructs the master equation from a list of objects | [24] |
| **GTPhMasterPixel** | constructs the master equation if the permittivity is given by a pixelmap | |
| **GTPhMPBBands** | imports a photonic band structure calculated by MPB | [36, 37] |
| **GTPhMPBFields** | imports electromagnetic field components calculated with MPB | [36, 37] |
| **GTPhPermittivity** | gives $\epsilon^{-1}(\vec{G})$ for all reciprocal lattice vectors used in the master equation | |
| **GTPhPixelSmooth** | gives a smoothed pixel map | |
| **GTPhPixelStructure** | allows to perform a modification of the permittivity distribution defined by a pixel map | |
| **GTPhPrismaticRod** | gives the structure factor of a prismatic rod | |
| **GTPhRodSmooth** | gives a smoothed structure factor of a rectangular rod | |
| **GTPhShowStructure** | plots an image of the arrangement of dielectric objects in a photonic structure | |
| **GTPhSlab** | gives the structure factor of a slab | |
| **GTPhSlabSmooth** | gives a smoothed structure factor of a slab | |
| **GTPhSphere** | gives the structure factor of a sphere | |
| **GTPhSymmetryBands** | performs a symmetry analysis of electromagnetic fields for a given band structure | [24] |
| **GTPhSymmetryField** | performs the symmetry analysis of an electromagnetic field | |
| **GTPhSymmetryPoint** | performs the symmetry analysis of electromagnetic fields given in datasets | |

## PseudoPotential.m

Table 10   Modules of the subpackage *PseudoPotential.m*. The package contains modules for the construction of plane-wave Hamiltonians.

| Name | Functionality | Ref. |
|---|---|---|
| **GTPwDatabaseInfo** | gives information about the pseudopotential parameter sets stored in a database | |
| **GTPwDatabaseRetrieve** | retrieves pseudopotential parameter sets stored in a database | |
| **GTPwDatabaseUpdate** | adds pseudopotential parameter sets to a database | |
| **GTPwDielectricF** | defines a screening function | |
| **GTPwEmptyLatticeIrep** | determines the irreducible representations of the empty lattice band structure | |
| **GTPwHamiltonian** | constructs a Hamiltonian based on pseudopotential theory | [44, 45] |
| **GTPwPrintParmSet** | prints a parameter set form a pseudopotential database | |

## RepresentationTheory.m

Table 11   Modules of the subpackage *RepresentationTheory.m*. The package contains modules for basic representation theory, like calculation of character tables and installation and handling of irreducible representations.

| Name | Functionality | Ref. |
|---|---|---|
| **GTAngularMomentumChars** | gives characters of a matrix representation in terms of Wigner $D$-matrices | |
| **GTAngularMomentumRep** | gives a matrix representation in terms of Wigner $D$-matrices | |
| **GTCharacterTable** | calculates the character table | [41] |
| **GTCharProjectionOperator** | applies the character projection operator to a given function | |
| **GTClebschGordanCoefficients** | calculation of Clebsch Gordan coefficients | [43] |
| **GTClebschGordanSum** | calculates the Clebsch-Gordan sum (direct sum) of two representation | |
| **GTClebschGordanTable** | illustrates calculated Clebsch-Gordan coefficients | |
| **GTDirectProductChars** | calculates characters of a direct product representation | |
| **GTDirectProductRep** | calculates matrices of a direct product representation | |
| **GTExtraRepresentations** | calculates the extra representations within a double group | |
| **GTGetIrep** | calculates matrices of irreducible representations | [42] |
| **GTIrep** | calculates the number of times an irred. rep. occurs within a red. rep. | |
| **GTIrepDimension** | calculates the dimensions of the irreducible representations | |
| **GTNumberOfIreps** | calculates the number of inequivalent irreducible representations | |
| **GTProjectionOperator** | applies the projection operator to a given function | |
| **GTReality** | estimates if a representation is potentially real, essentially complex or pseudo-real | |
| **GTRegularRepresentation** | calculates the regular representation of a group | |
| **GTSOCSplitting** | calculates the splitting of states due to spin-orbit coupling | |
| **GTSpinCharacters** | calculates the characters of the spin representation | |
| **GTWignerProjectionOperator** | applies the projection operator to spherical harmonics | |

**TightBinding.m**

Table 12 Modules of the subpackage *TightBinding.m*. The package contains modules for the construction of tight-binding Hamiltonians in two- and three-center form.

| Name | Functionality | Ref. |
|---|---|---|
| **GTFindStateNumbers** | findes the number of states within a certain energy interval (real space) | |
| **GTHamiltonianPlot** | illustrates zero and non-zero entries within the tight-binding Hamiltonian | |
| **GTPlotStateWeights** | plots the contribution of atomic sites to an electronic state (real space) | |
| **GTSymmetryBasisFunctions** | calculates the associated irreducible representation of a wave function | |
| **GTTbAtomicWaveFunction** | calculates values of an atomic-like wave function | |
| **GTTbDatabaseInfo** | gives information about tight-binding parameters stored in a database | [63] |
| **GTTbDatabaseRetrieve** | retrieves tight-binding parameters from a database | |
| **GTTbDatabaseUpdate** | adds tight-binding parameters to a database | |
| **GTTbGetParameter** | gives a particular tight-binding parameter from a database | |
| **GTTbHamiltonian** | constructs a tight-binding Hamiltonian | [46] |
| **GTTbHamiltonianElement** | constructs a single element of a tight-binding Hamiltonian | |
| **GTTbHamiltonianRS** | constructs a real space tight-binding Hamiltonian | |
| **GTTbNumberOfIntegrals** | calculates the number of independent tight-binding integrals | [47] |
| **GTTbParameterNames** | creates a set of parameter names | |
| **GTTbParmToRule** | gives a rule to replace tight-binding symbols using a given parameter set | |
| **GTTbPrintParmSet** | prints a tight-binding parameter set from a database | |
| **GTTbRealSpaceMatrix** | constructs the interaction of two atoms in a tight-binding Hamiltonian | |
| **GTTbSpinMatrix** | gives elementary spin matrices for tight-binding Hamiltonians | [61] |

| | |
|---|---|
| **GTTbSpinOrbit** | adds spin-orbit coupling to a given tight-binding Hamiltonian |
| **GTTbSymbol2C** | gives a symbol according to the nomenclature of the two-center approximation |
| **GTTbSymmetryBandStructure** | performs a symmetry analysis of a given band structure |
| **GTTbSymmetryPoint** | performs a symmetry analysis of energy bands at a specified $\vec{k}$-point |
| **GTTbSymmetrySingleBand** | performs a symmetry analysis of a single energy band |
| **GTTbToFortran** | transforms a k-space tight-binding Hamiltonian into a FORTRAN module |
| **GTTbToFortranList** | prints a Hamiltonian as FORTRAN code |
| **GTTbWaveFunction** | constructs a tight-binding wave-function |

## Vibrations.m

Table 13 Modules of the subpackage *Vibrations.m*. The package contains modules for the study of vibrational modes of solids and molecules.

| Name | Functionality | Ref. |
|---|---|---|
| **GTVibDisplacementRep** | gives the displacement representation of a molecule | |
| **GTVibDynamicalMatrix** | gives the dynamical matrix for a given structure | |
| **GTVibModeSymmetry** | gives the vibrational modes of a molecule | |
| **GTVibSetParameters** | substitutes spring constants and masses within a dynamical matrix | |
| **GTVibTbToPhonon** | transforms a tight-binding p-Hamiltonian into a dynamical matrix | [64] |
| **GTVibTbToPhononRule** | gives rules to transform a tight-pinding p-Hamiltonian into a dynamical matrix | |